\documentclass[showpacs,twocolumn,floatfix,amsmath,amssymb,aps,pra]{revtex4}
\usepackage{graphicx}
\usepackage{latexsym,epsfig,bm,times,psfrag,subfigure}
\usepackage{bm,times}
\usepackage{dcolumn}
\begin{document}

\newcommand{\tiz}{\tilde{z}}
\newcommand{\tit}{\tilde{t}}
\newcommand{\tix}{\tilde{x}}
\newcommand{\tiy}{\tilde{y}}

\title{Caustic formation in expanding condensates of cold atoms}
\author{J.\ T.\ Chalker$^{1}$ and B. Shapiro$^{1,2}$}
\affiliation{$^{1}$Theoretical Physics, Oxford University, 1, Keble Road, Oxford, OX1 3NP, United Kingdom\\}
\affiliation{$^{2}$Department of Physics, Technion - Israel Institute of Technology, Haifa 32000, Israel}
\date{\today}
\begin{abstract}
We study the evolution of density in an expanding Bose-Einstein
condensate that initially has a spatially varying phase, concentrating
on behaviour when these phase variations are large. In this regime
large density fluctuations develop during expansion. Maxima have a
characteristic density that diverges with the amplitude of phase
variations and their formation is analogous to that of caustics in
geometrical optics. We analyse in detail caustic formation in a
quasi-one dimensional condensate, which before expansion is subject to
a periodic or random optical potential, and we discuss the equivalent
problem for a quasi-two dimensional system. We also examine the influence
of many-body correlations in the initial state on caustic formation
for a Bose gas expanding from a strictly one-dimensional trap. In
additon, we study a similar arrangement for non-interacting fermions,
showing that Fermi surface discontinuities in the momentum
distribution give rise in that case to sharp peaks in the spatial
derivative of the density. We discuss recent experiments and argue
that fringes reported in time of flight images by Chen and co-workers
[Phys. Rev. A {\bf 77}, 033632 (2008)] are an example of caustic
formation.
\end{abstract}
\maketitle

\section{Introduction}
\label{introduction}
Focussing of rays and the associated phenomenon of caustic formation
are both well known in optics \cite{intro}. For the phase screen model
caustics have been studied extensively by Berry \cite{berry}. In this
model a monochromatic plane wave encounters a thin screen, located on
the plane $z{=}0$ and having coordinates $\zeta$ and $\eta$ within the
plane. The screen impresses on the wave a phase $\theta(\zeta,\eta)$,
which may be deterministic or random. In either case, for strong
variation of this phase, a wave propagating in the $z$-direction and
passing through the screen will develop large intensity
variations. Specifically, observation of the wave intensity at a point
sufficiently far beyond the screen will reveal a pattern of bright
lines. These are caustics. Within geometrical optics, light intensity
on caustics diverges. Diffraction effects smooth these singularities
and decorate caustics with interference fringes.

A similar phenomenon can occur with matter waves associated with
propagating clouds of cold atoms, especially if these form a
Bose-Einstein condensate (BEC). The purpose of this paper is to
develop a theory of caustics for cold atoms and to discuss the
experimental conditions for their observation. The arrangement we
consider differs substantially from that in optics. Caustics develop
during the expansion of an atomic cloud released from a trap. The
corresponding matter wave is not at all monochromatic, and time
assumes the role of the spatial axis of propagation in optics. The
mechanism of impression of the phase is also different. One
possibility is to create  density variation in the trap. During the
expansion this initial density modulation, in combination with strong
non-linearity, produces a space dependent velocity field, thus
impressing a phase on the BEC \cite{clement}. In Sec. \ref{q1d} we
shall discuss further this mechanism, and the way it can lead to
formation of caustics, for quasi-one dimensional systems. Another
possibility for impressing phase variation is by applying to a trapped
condensate a short pulse of a space-dependent potential. Immediately
after the pulse, the trapping potential is switched off so that the
condensate starts its free expansion  with the phase variation
generated by the pulse. We employ this mechanism in Sec. \ref{s1d},
where we discuss the strictly one-dimensional case.

In both examples, the geometrical optics limit is the regime in which
the impressed phase has spatial variations that are much larger than
unity. Characterising these phase variations by an amplitude
$\theta_0$ and a spatial scale $R_0$, our central conclusion is that
for $\theta_0 \gtrsim 1$ an expanding condensate develops caustics
after an expansion  time of order $t^*  = m R_0^2/\hbar \theta_0$,
where $m$ is the atomic mass. The density close to caustics diverges
with $\theta_0$, as $\theta_0^{1/3}$ in the simplest case, and may
therefore be much larger than it is in the background between
caustics. 

The development of density fluctuations as a consequence of initial
phase fluctuations has been studied in previous experimental and
theoretical work, both for the case when
these phase fluctuations are thermal in origin \cite{dettmer}, and 
for the case in which they arise from a disordered potential applied
to the trap\cite{clement}. In this earlier work, however, the
significance of $\theta_0$ was not identified, and the method used were applicable for $\theta_0
\ll 1$, when density fluctuations never become large.

In outline, the organisation and main results of this paper are as
follows. In Sec. \ref{q1d} we treat quasi-one dimensional systems with
an initial density modulation, using the Gross-Pitaevskii equation to
describe the conversion of density to phase modulation. Relative
density variations generated in a trap by a potential with amplitude
$V_0$ are small if $V_0$ is much smaller than $\mu$, the chemical
potential. Nevertheless, they may lead to phase fluctuations that are
large, since with radial frequency $\omega_\perp$ we find $\theta_0
\sim V_0/\hbar \omega_\perp$. We suggest that the large density
contrast measured recently \cite{chen} for a condensate expanding from
a trap with a disordered optical potential  should be understood as an
example of caustic formation. We also examine behaviour as $V_0$ is
increased to values larger than $\mu$, inducing fragmentation of the
condensate. We show that caustics are not formed in the expansion of a
highly fragmented condensate, but find that the value of $V_0$ above
which they are eliminated is larger than the threshold for
fragmentation. In Sec. \ref{s1d} we examine the effect of many-body
correlations in the initial wavefunction on caustic formation, taking
these correlations from the Lieb-Liniger model. We find that behaviour
is controlled by the value of the healing length $\xi$: taking
$\theta_0 \gg 1$,  caustics survive if interactions are weak and
$\theta_0^{1/3} \xi/R_0 \gg 1$,
while in the opposite limit they are suppressed.  We also consider, in
Sec. \ref{2d}, expansion of a condensate from a two-dimensional trap
in which there is a smooth, spatially varying potential. In this
geometry caustics form a network of intersecting curves for $\hbar
\omega_\perp \ll V_0 \ll \mu$. Segments of these curves are eliminated
for $V_0  \sim \mu$, although without any sharp signature of the
percolation transition that takes place at $\mu = \mu_{\rm c}$ for the condensate in the
trap. For $\mu \ll \mu_{\rm c}$  caustic formation is suppressed altogether,
as in quasi-one dimension. We discuss briefly behaviour for fermion
systems in Sec.~\ref{fermions}, and close with a summary in Sec. \ref{summary}.

\section{Quasi-one dimension: mean field theory}
\label{q1d}

In this section we consider a strongly anisotropic BEC, initially
confined in the radial direction by a harmonic trap with frequency
$\omega_\perp$. In experiments there is also weak confinement in the
axial direction, with frequency $\omega_z \ll \omega_\perp$. The axial
confinement will be neglected, which implies that our considerations
are limited to times $t \ll 1/\omega_z$ after the release of the
condensate from the trap. We use $z$ and $\rho$ as axial and radial
coordinates.

Prior to its release the condensate is in its ground state (we assume
zero temperature) in the presence of a radial confining potential
$\frac{1}{2}m \omega_\perp^2\rho^2$  and a $z$-dependent potential
$V(z)$. In order to clarify the mechanism of caustic formation we
shall start by treating a periodic potential $V(z)= V_0 \cos k_0 z$,
and then proceed to the case of a disordered potential. In the latter
example we use $V_0$ to denote the characteristic amplitude of the
random potential, and $R_0$ its correlation length, with $R_0 \sim
1/k_0$ for the two cases to be comparable. We assume a smoothly
varying potential, in the sense that $k_0 a_\perp \ll 1$, where
$a_\perp = \sqrt{2\mu/m\omega_\perp^2}$ is the radius of the BEC in
the trap, with $\mu$ the chemical potential, assumed much larger than
$\hbar \omega_\perp$.

The potential $V(z)$ produces density modulations of the BEC in the
trap. Within the Thomas-Fermi approximation the ground state density
is
\begin{equation}
n_0(\rho,z) = \frac{1}{g}\left(\mu -V(z) - \frac{1}{2}m \omega_\perp^2 \rho^2 \right)\;,
\label{initial}
\end{equation}
where $g$ is the coupling constant for the non-linear term in the Gross-Pitaevskii equation.

At time $t{=}0$ all potentials are switched off and the condensate expands according to the equations \cite{pitaevskii}
\begin{eqnarray}
\frac{\partial n}{\partial t} &+& {\rm div}\,n \vec{v} = 0\\
m\frac{\partial \vec{v}}{\partial t} &+& \vec{\nabla} \left(\frac{1}{2}mv^2 + gn \right) = 0\label{z-evolution}
\end{eqnarray}
where $\vec{v} = (v_\rho,v_z)$ is the condensate velocity, related to
its phase $\theta$ by $\vec{v} = \frac{\hbar}{m} \vec{\nabla}
\theta$. These equations are to be solved with the initial condition
Eq.~(\ref{initial}), supplemented by the requirement that the velocity
field $\vec{v} = 0$, or equivalently that the phase is uniform, at the
start of the expansion.

The condensate undergoes rapid radial expansion, according to the
standard scaling picture \cite{pitaevskii}, but due to the initial
density modulation it also develops an axial velocity component
$v_z(z,t)$. The latter is governed by the $z$-component of
Eq.~(\ref{z-evolution}) . During the initial stage of radial expansion
we may neglect the kinetic energy $\frac{1}{2}mv^2$ compared to the
interaction energy $gn$, and obtain \cite{clement} 
\begin{equation}
v_z(z,t) = \frac{1}{m\omega_\perp}\frac{{\rm d} V(z)}{{\rm d} z} \arctan \omega_\perp t \;.
\end{equation}
This corresponds to an impressed phase
\begin{equation}
\theta(z) = \frac{\pi}{2\hbar \omega_\perp} V(z)
\label{theta}
\end{equation}
at any time $t_0$ lying in the window $t^* \gg t_0 \gg 1/\omega_\perp$.
Both this phase and the consequences of non-zero $v_z(z,t_0)$ for the
subsequent time evolution have been discussed in Ref.~\cite{clement},
but the theory developed there applies only to systems with
sufficiently weak initial density modulations. In the following we
show that the typical phase magnitude $\theta_0 \sim V_0/\hbar
\omega_\perp$ is the relevant parameter: for $\theta_0 \ll 1$ only
weak density modulations develop at later times, but for $\theta_0
\sim 1$ density modulations arise that are comparable to the average
density, while in the limit $\theta_0 \gg 1$ caustics appear. Large
effects of this kind have been observed in a recent experiment
\cite{chen}.  As we shall see, they develop at characteristic times of
order $t^{*} = m/\hbar k_0^2 \theta_0$ which is parametrically larger
than $1/\omega_\perp$ by the factor $\mu/V_0 k_0^2 a_\perp^2$.

During the second stage of expansion, at times much larger than $t_0$,
the nonlinearity of the Gross-Pitaevskii equation can be neglected and
we arrive at the problem of linear time evolution of the BEC
wavefunction with, as an initial condition, the impressed phase
$\theta(z)$ given by Eq.~(\ref{theta}). The wavefunction can be
factorised into radial and axial parts, as
\begin{equation}
\Psi(\rho,z,t) = \Phi(\rho,t)\psi(z,t)
\end{equation}
and the density is
\begin{equation}
n(\rho,z)= |\Phi(\rho,t)|^2 |\psi(z,t)|^2\;.
\end{equation}
The behaviour of $|\Phi(\rho,t)|^2$ is well established and given by
the radial scaling function of Ref.~\cite{pitaevskii}. Our concern in
the following is with the axial part, $|\psi(z,t)|^2$, which gives the
density at point $z$ and time $t$, normalised by the radial factor
$|\Phi(\rho,t)|^2$. The function $\psi(z,t)$ satisfies the linear
Schr\"odinger equation
\begin{equation}
{{\rm i}}{\hbar} \frac{\partial \psi}{\partial t} = - \frac{\hbar^2}{2m} \frac{\partial^2 \psi}{\partial z^2}
\label{schr}
\end{equation}
with the initial condition
\begin{equation}
\psi(z,t_0) = {\rm e}^{{\rm i} \theta(z)}\;.
\label{init}
\end{equation}
This form for $\psi(z,t_0)$ neglects the initial modulation of the BEC
density, which is justified if $V_0 \ll \mu$ so that the relative
variation at time $t=t_0$ is much smaller than unity. Since we are
interested in density variations, at later times, that are  of order
unity or larger, we neglect these small initial modulations, except
for their crucial part in producing the phase $\theta(z)$. For most of
this section we impose the condition $V_0 \ll \mu$ (but not the more
stringent one $V_0 \ll \hbar \omega_\perp$ implicitly assumed in
Ref.~\cite{clement}). The opposite case $V_0 > \mu$ corresponds to a
fragmented BEC and will be treated separately towards the end of the
section.

The solution to Eqns~(\ref{schr}) and (\ref{init}) for $t\gg t_0$ is
\begin{equation}
\psi(z,t) =  \sqrt{\frac{m}{2\pi {\rm i} \hbar t}}   \int {\rm
  d}z'\;\exp\left[\frac{{\rm i}m}{2\hbar t}(z-z')^2+{\rm i}
  \theta(z')\right]\;,
\label{psi-1}
\end{equation}
where the impressed phase $\theta(z)$ can be an arbitrary function of $z$.

Let us now consider $\theta(z) = \theta_0\cos k_0 z$ and introduce the
dimensionless variables $\tiz = k_0 z$ and $\tit = t/t^* = \hbar
k_0^2 \theta_0 t/m$, so that
\begin{equation}
\psi(\tiz , \tit) = \sqrt{\frac{\theta_0}{2\pi {\rm i} \tit}} \int
{\rm d} \zeta \;\exp\left[{\rm i} \theta_0
  \varphi(\zeta,\tiz,\tit)\right]\;,
\label{psi}
\end{equation}
with
\begin{equation}
\varphi(\zeta,\tiz,\tit) = \frac{(\tiz-\zeta)^2}{2 \tit} + \cos \zeta \;.
\end{equation}
The relative density $|\psi(z,t)|^2$, initially unity, acquires
spatial variations with the passage of time. For $\theta_0 \ll 1$
relative density variations remain small for all times, as is clear
from expanding  the factor $\exp({\rm i} \theta_0 \cos \zeta)$ in the
integrand of Eq~(\ref{psi}). This is the regime considered for a
random potential in Ref.~\cite{clement}. We study the opposite case,
$\theta_0 \gg 1$, when the relative density can develop large
modulations and caustics can be formed. This is the regime encountered
in the experiment of Ref.~\cite{chen}.

\begin{figure}
\includegraphics[scale=0.38]{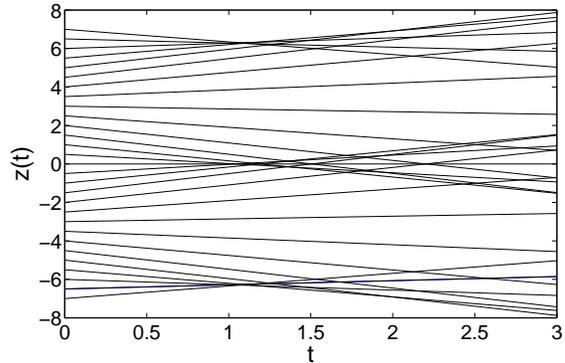}
\caption{Paths $z(t)$ followed within the geometrical optics
  approximation by atoms expanding from a condensate with an initial
  phase $\theta(z) = \theta_0 \cos(z)$.} \label{fig1}
\end{figure}

We will be interested in the form of the relative density at a given
instant $t$. This gives the $z$-dependence of the density of an
expanding atomic cloud, measured with a probe beam perpendicular to
the condensate axis. For $\tit \ll 1$, relative density modulations
are small. They grow linearly with $\tit$ and are oscillatory in
$\tiz$ with period $2\pi$. Growth of density maxima with time
culminates, for $\theta_0 \gg 1$, in the formation of caustics at
times $\tit \gtrsim 1$.  In this regime Eq.~(\ref{psi}) can be
evaluated by the stationary phase method. Caustics are determined by
the vanishing of not only the first but also the second derivative of
the phase $\varphi(\zeta,\tiz,\tit)$ with respect to $\zeta$:
\begin{eqnarray}
\frac{\partial \varphi(\zeta,\tiz,\tit)}{\partial \zeta} &=& \frac{\zeta - \tiz}{\tit} - \sin \zeta = 0 \label{rays}\\
\frac{\partial^2 \varphi(\zeta,\tiz,\tit)}{\partial \zeta^2} &=&\frac{1}{\tit} - \cos \zeta = 0\;. \label{singular}
\end{eqnarray}
These equations have a simple geometric interpretation, in complete
analogy with optics \cite{berry,hannay}, which is illustrated in Fig.~\ref{fig1}. Eq~(\ref{rays}) defines rays
of atoms, in the sense that  atoms emerging from a point $\zeta$ have
a velocity $-\sin \zeta$ given by the derivative of the phase $\cos
\zeta$, and so reach the point $\tiz = \zeta - \tit \sin \zeta$ at
time $\tit$. The roots of Eq.~(\ref{singular}) identify singular
points $\zeta_n$. Vanishing of $\partial^2_\zeta
\varphi(\zeta,\tiz,\tit)$ at these points means that the emerging rays
are focussed on the observation points. Thus, for a given $\tit$, one
will observe caustics, as bright spots, at points $\tiz_n = \zeta_n -
\tit \sin \zeta_n$. For $\tit \gg 1$, $\tiz_n \approx \frac{\pi}{2} +
n \pi - (-1)^n \tit$ with $n=0\,, \pm 1\,, \ldots$. The density at
these points is found by computing the integral in
Eq.~(\ref{psi}). Within the stationary phase method, since at the
singular points $\tiz=\tiz_n$ the first two derivatives of
$\varphi(\zeta,\tiz,\tit)$ vanish, the integral is controlled by the
third  derivative, $\partial^3_\zeta \varphi(\zeta,\tiz,\tit)$, and
has a value proportional to $\theta_0^{-1/3}$, resulting in a large
relative density $|\psi(\tiz,\tit)|^2 \sim \theta_0^{1/3}$. For times
$\tit \gg 1$ the density in the vicinity of caustics decays as
$\tit^{-1}$, as is clear from the prefactor to the integral in
Eq.~(\ref{psi-1}).

At caustics one can observe spectacular diffraction effects
\cite{berry}. On slight deviation $\delta \tiz \sim \theta_0^{-2/3}$
from the point $\tiz_n$ the density displays a sharp drop, followed by
aperiodic oscillations. The detailed shape of these oscillations can
be found by further studying the integral in Eq.~(\ref{psi}). We shall
not do this here, but rather remark on the picture between caustics,
at a point $\tiz$  well separated from all $ \tiz_{n}$. In this case
only Eq~(\ref{rays}) remains to be satisfied, and since it can have
several solutions, several saddle points will contribute to the
integral in Eq.~(\ref{psi}). Each saddle point contributes a term of
order $\theta_0^{-1/2}$ which cancels the prefactor of
$\theta_0^{1/2}$. Interference between different contributions ---
that is, between different rays arriving at the point $\tiz$ at time
$\tit$ --- results in a density pattern with variations of order unity
and spacing between fringes of order $\Delta \tiz \sim 1/\theta_0 $.

Our discussion has been limited to caustics of the simplest kind,
referred to as {\it folds} in Ref.~\cite{berry}. More singular
caustics, known as {\it cusps} can also occur. These stem from rays
emerging from the points $\zeta_n = 2 \pi n$ with $n=0\,,\pm1\,,
\ldots$ and are visible only at a patricular time instant, $\tit
=1$. The point is that for $\zeta_n = 2 \pi n$ and $\tit =1$ not only
the first two but also the third derivative, $\partial^3_\zeta
\varphi(\zeta,\tiz,\tit)= \sin \zeta_n$, vanishes. The integral in
Eq.~(\ref{psi}) is then controlled by the fourth derivative,
$\partial^4_\zeta \varphi(\zeta,\tiz,\tit)= 1$, and is of order
$\theta_0^{-1/4}$, which implies that the reduced density at the
points $\tiz_n=2\pi n$ is at this time $|\psi(\zeta,\tit)|^2 \sim
\theta_0^{1/2}$. Studying the integral in Eq.~(\ref{psi}) in more
detail, for values of $\tit$ and $\tiz$ near $\tit=1$ and $\tiz = 2
\pi n$ one can identify lines in the $\tiz$-$\tit$ plane on which
$|\psi(\zeta,\tit)|^2$ drops from its maximum value, of order
$\theta_0^{1/2}$, to values of order $\theta_0^{1/3}$. These lines
have cusps at the most singular points, located at $\tiz=2\pi n$ and
$\tit=1$.

We illustrate these ideas using a numerical evaluation of
Eq.~(\ref{psi}) to obtain the relative density $|\psi(z,t)|^2$. The
results are shown for several values of $\tit$ at fixed $\theta_0$ in
Fig.~2, and for several values of $\theta_0$ at fixed $\tit$ in
Fig.~3. In the large $\theta_0$ limit, caustics are located at:
$\tiz=0$ for $\tit
= 1$ at; $\tiz= \pm 0.685 $ for $\tit =2$; and at $\tiz = \pm 1.598$
for $\tit=3$. The two figures show that $|\psi(z,t)|^2$ has peaks near
these points, which grow in amplitude and become increasingly well defined
with larger $\theta_0$.

\begin{figure}
\includegraphics[scale=0.25]{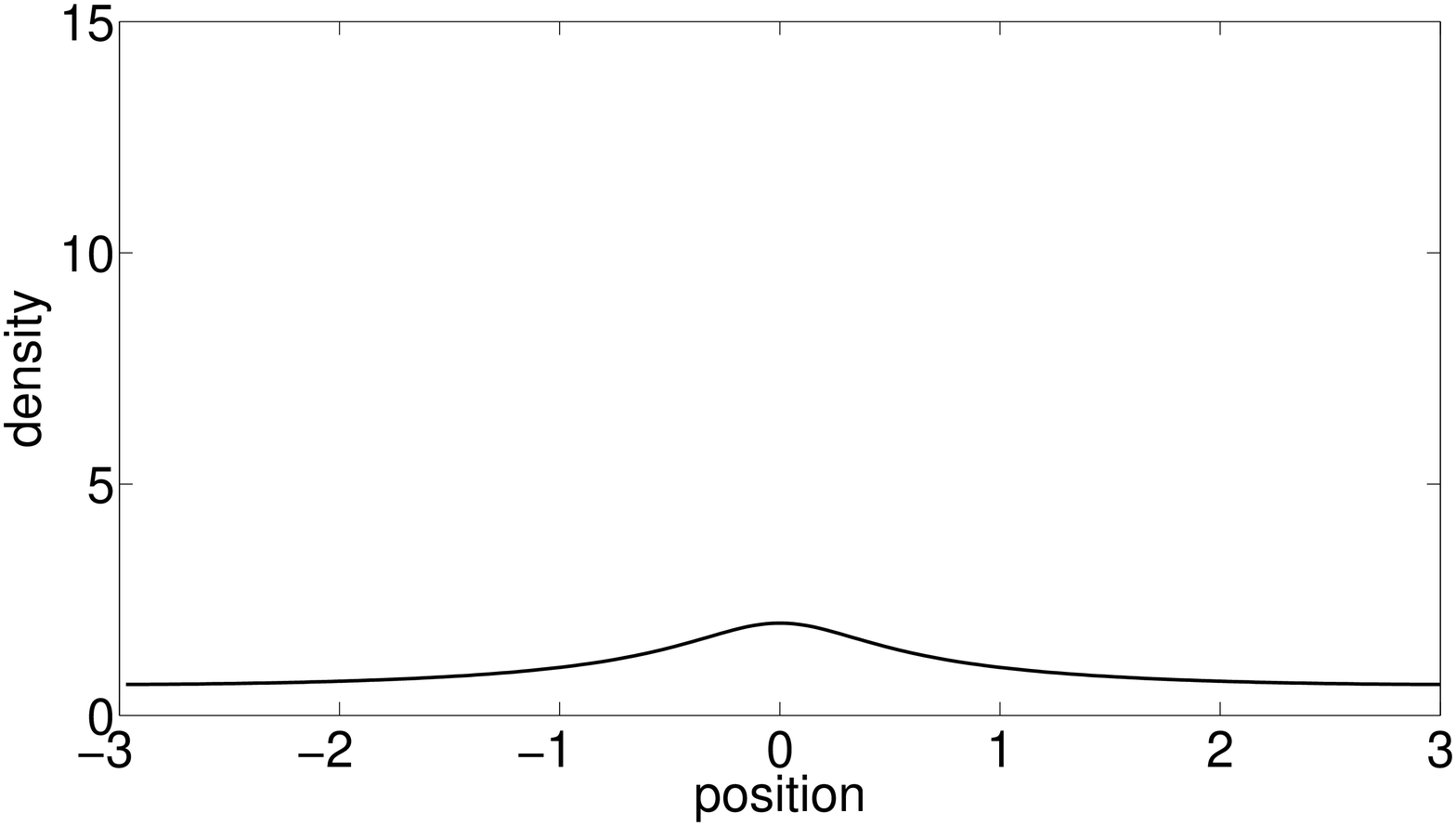}
\includegraphics[scale=0.25]{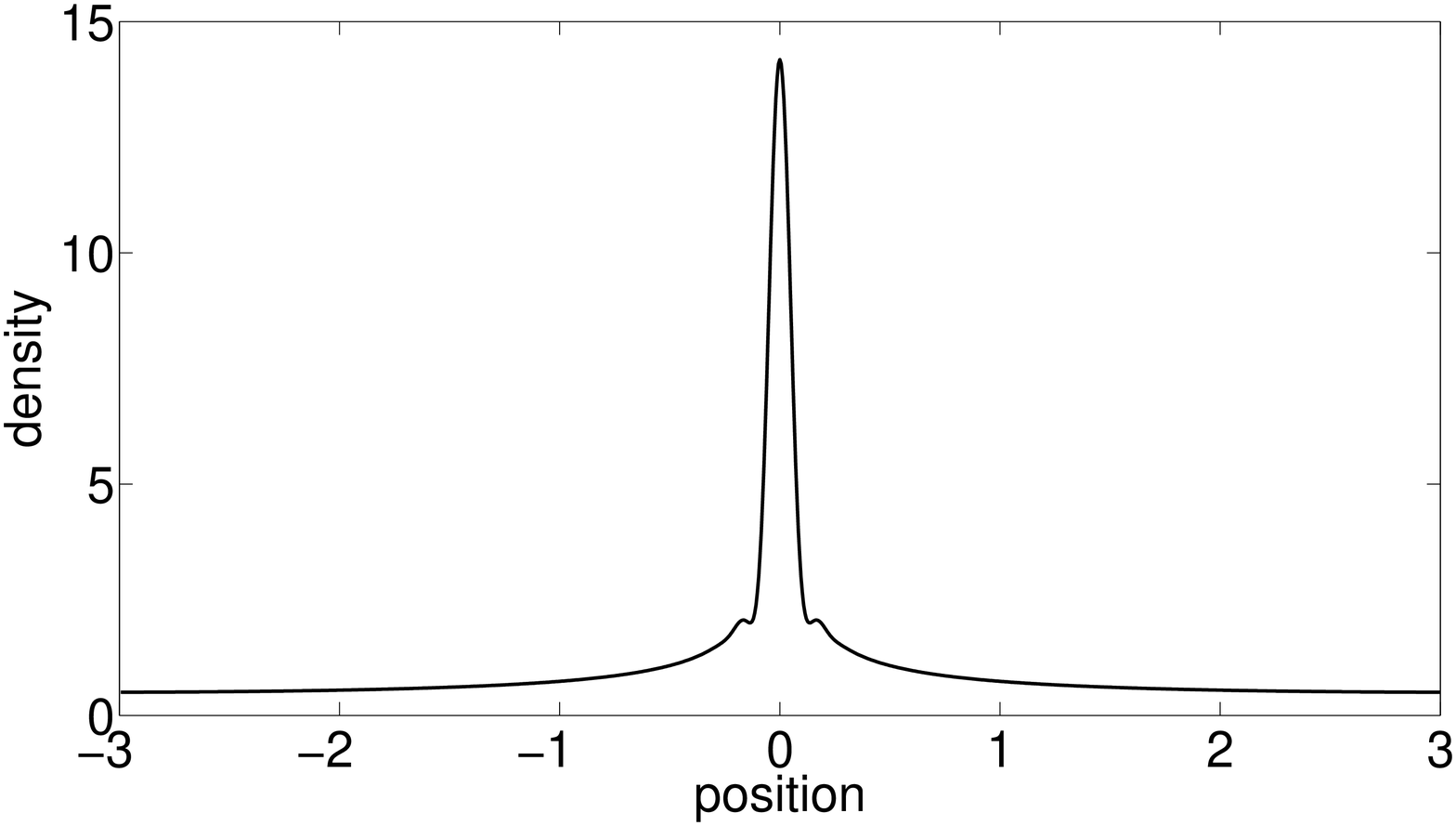}
\includegraphics[scale=0.25]{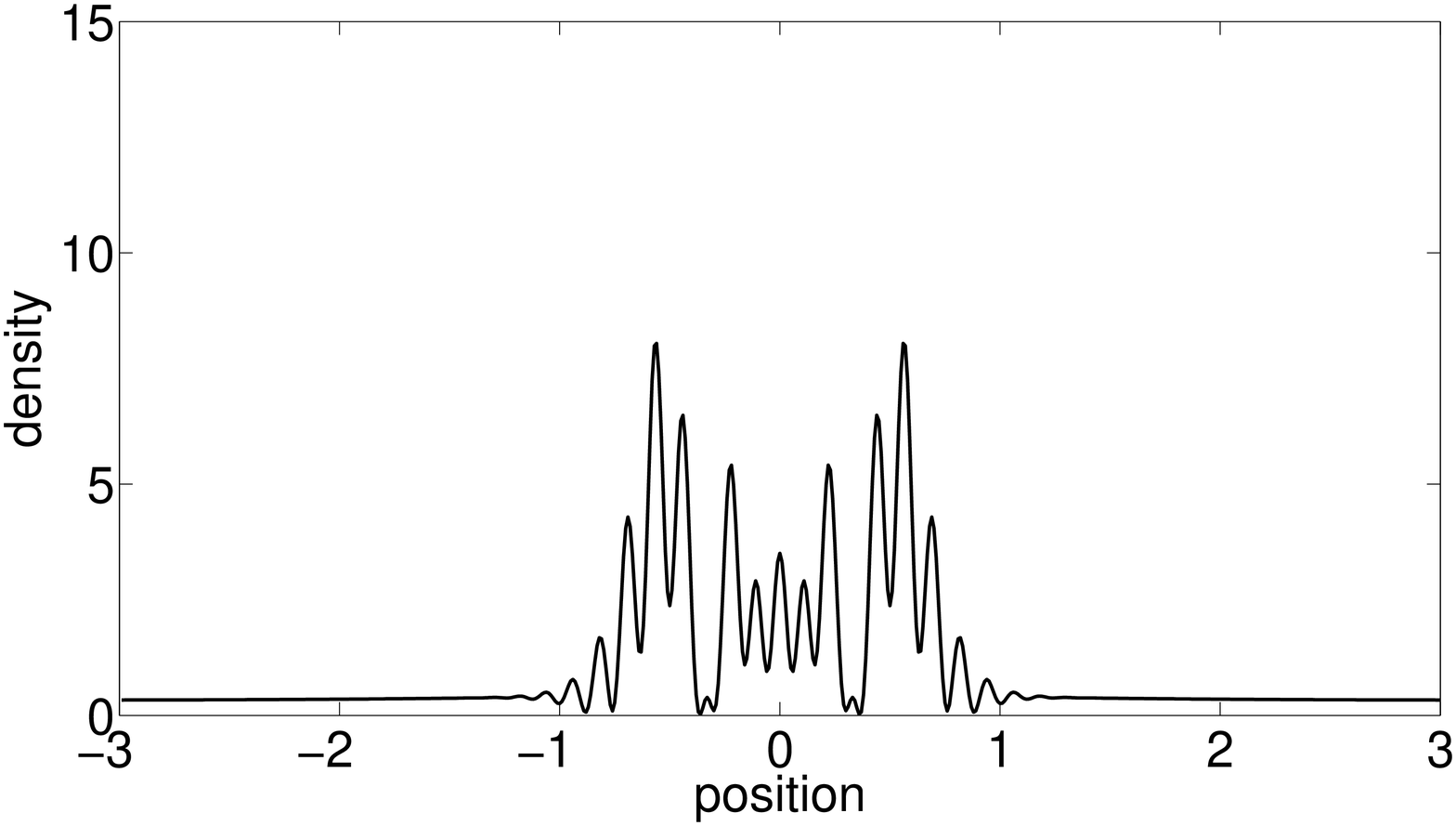}
\caption{Relative density $|\psi(\tiz,\tit)|^2$ as a function of position
  $\tiz$, for $\theta_0=30$ and $\tit = 0.5$ (top), $1$ (middle), and
  $2$ (bottom).} 
\end{figure}

\begin{figure}
\includegraphics[scale=0.25]{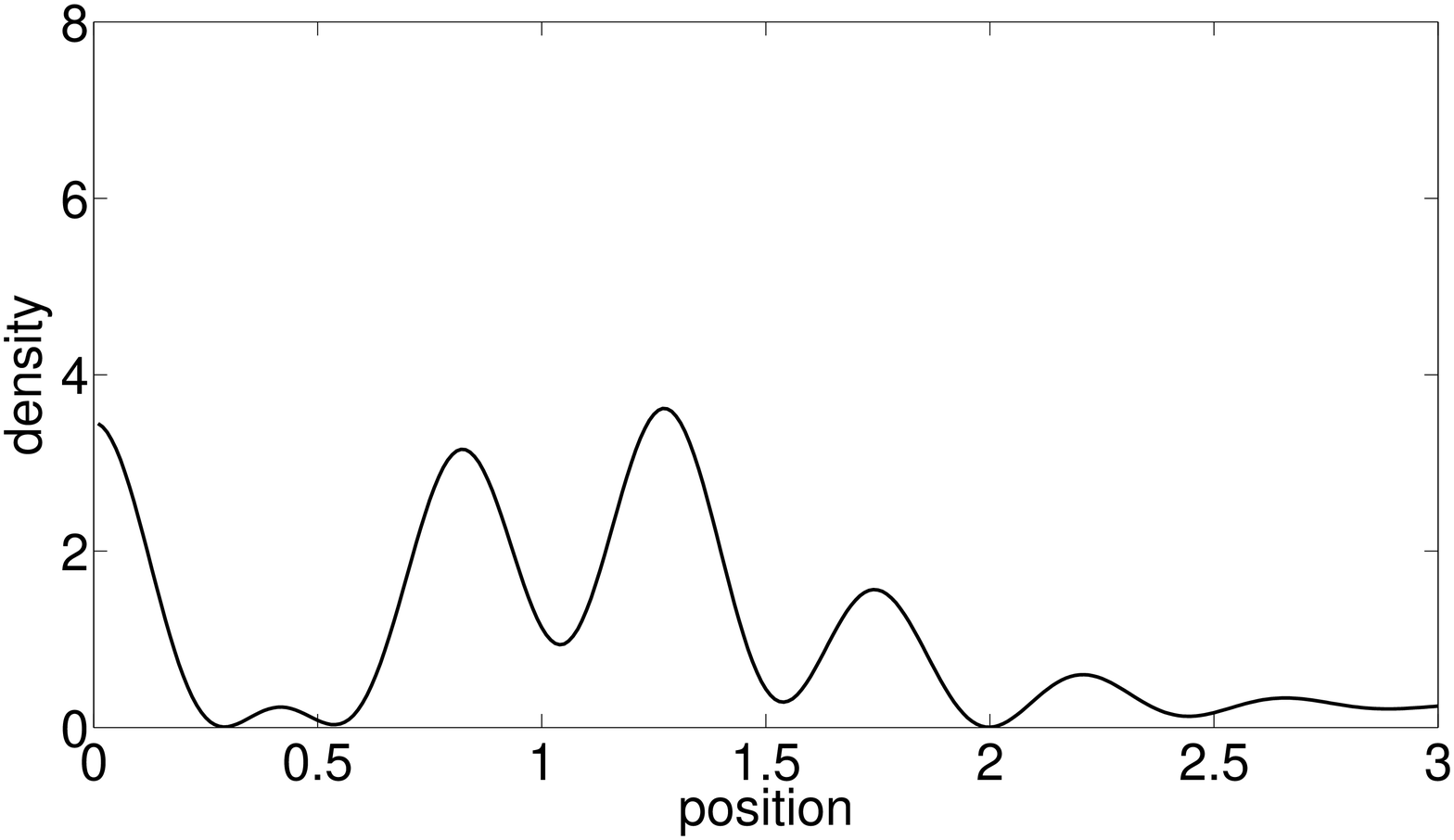}
\includegraphics[scale=0.25]{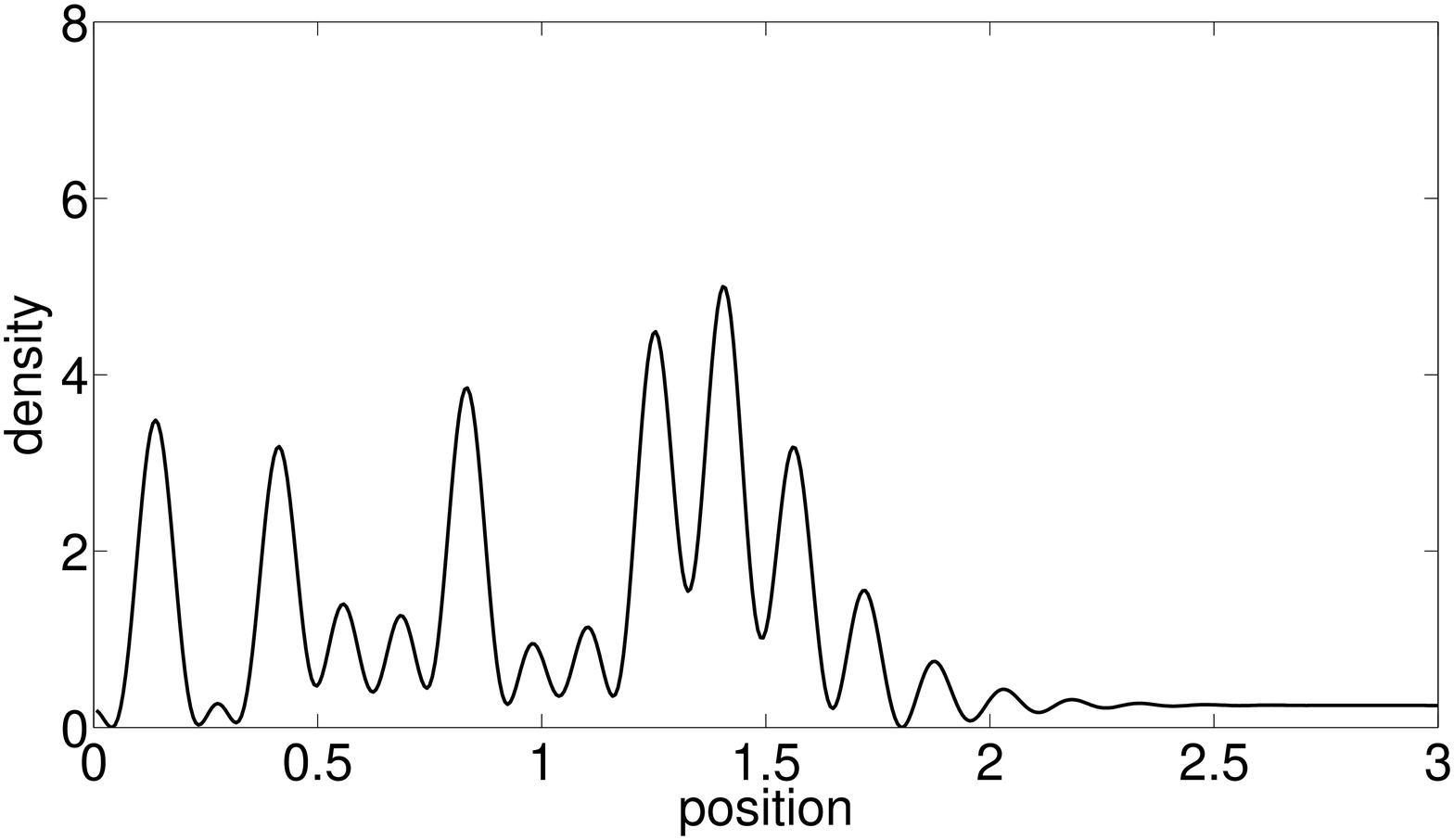}
\includegraphics[scale=0.25]{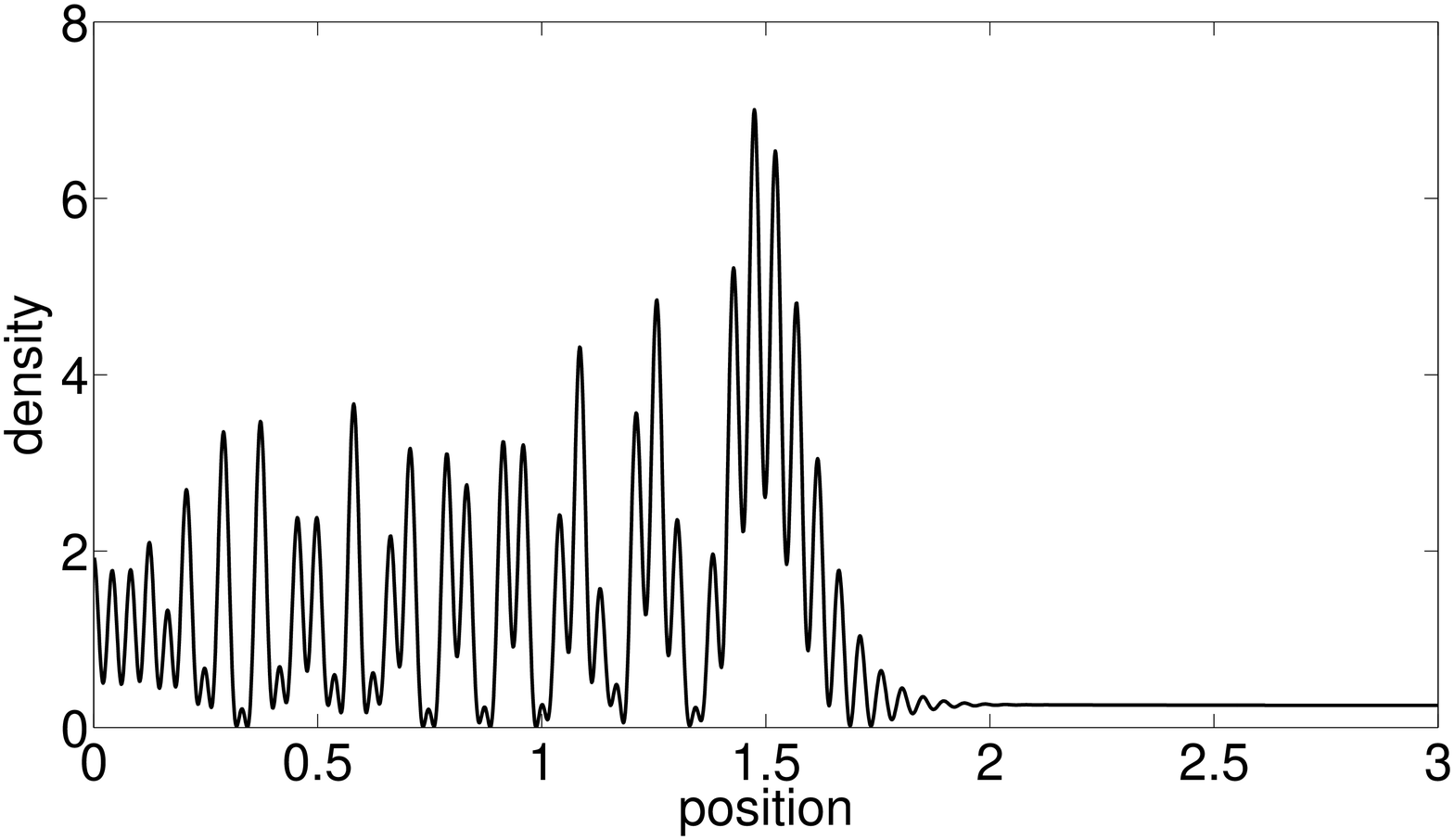}
\caption{Relative density $|\psi(\tiz,\tit)|^2$ as a function of position
  $\tiz$, for $\tit=3$ and $\theta_0 = 10$ (top), $30$ (middle), and
  $100$ (bottom).} 
\end{figure}

We now turn to a disordered potential $V(z)$, which we take to be
Gaussian distributed with zero mean, amplitude $V_0$, correlation
length $R_0$, and correlation function
\begin{equation}
\overline{ V(z)V(z')} = V_0^2 f\left(\frac{z-z'}{R_0}\right)\;.
\label{disorder-correlator}
\end{equation}
Here and elsewhere, we use an overbar
$\overline{\phantom{(}\dots\phantom{)}}$ to denote a disorder
average. The function  $f(\tiz)$ has unit amplitude and unit range, so
that $f(0)=1$ and $f(\tiz) \to 0$ for $\tiz \gg 1$.  This implies for
the impressed phase that its mean is zero and
\begin{equation}
\overline{\theta(z)\theta(z')} = \theta_0^2 f\left(\frac{z-z'}{R_0}\right)
\end{equation}
with $\theta_0 = \pi V_0/2\hbar \omega_\perp$. The experimentally
realised disordered potentials for cold atoms are optical speckle
potentials (see Ref.~\cite{kuhn} for an extensive discussion) whose
correlation function is
\begin{equation}
f(\tiz) = \frac{\sin^2(\tiz)}{\tiz^2}\;.
\end{equation}
Eq.~(\ref{psi-1}) for $\psi(z,t)$ applies also for a disordered
potential and the mechanism for caustic formation is qualitatively the
same as for the harmonic potential discussed above. Caustics are due
to rays emerging from singular points $z'_n$ on which there are zeros
of two or more derivatives with respect to $z'$ of the overall phase
in Eq.~(\ref{psi-1}),
$$
\frac{m}{2\hbar t}(z-z')^2+ \theta(z')\;.
$$
Since the impressed phase $\theta(z')$ is now a random function, the
formation of caustics of different types (fold, cusp or more singular)
is a matter of probability. The typical time for caustic formation is
$t^*=mR_0^2/\hbar \theta_0 $. Using the relations
$\frac{1}{2}m \omega_\perp^2 a_\perp^2 = \mu$ and $\theta_0 = \pi
V_0/2\hbar \omega_\perp$, the formation time can be written in terms
of the experimentally controlled parameters as
$$
t^* = \frac{4}{\pi \omega_\perp}\frac{\mu}{V_0}\left(\frac{R_0}{a_\perp}\right)^2\;.
$$
It is quite straightforward to modify the theory of optical caustics
\cite{berry,hannay} to the case of caustics in an expanding BEC, but
we do not pursue this further here.

So far our discussion of caustics for the random case was qualitative
and pertained to a specific, typical realisation of the disordered
potential. We next present some quantitative analytic results for the
second moment of the reduced density fluctuations,
$$
\overline{|\psi(z,t)|^4} \equiv  S(t) +1\;.
$$
The averaging restores translational invariance and so eliminates the
$z$-dependence. We denote the resulting (time-dependent) quantity by
$S(t) + 1$ in order to emphasise the analogy with speckle patterns in
optics. There $S$ is called the scintillation index and it is a
measure of spatial intensity fluctuations in the speckle pattern. For
a uniform intensity $S$ is clearly zero, and for the standard speckle
pattern, created by  great number of interfering waves with random
phases, in which the intensity has a Rayleigh distribution,  $S$ is
unity \cite{goodman}. For the random screen problem, $S$ was studied
as a function of distance $D$ from the screen in
Ref.~\cite{jakeman}. It was shown there that, at first, $S$ increases
with $D$ and reaches a maximum value larger that $1$ but then, with
further increase in $D$, the value of $S$ drops and approaches $1$ for
large $D$. We briefly outline a similar calculation for our problem.

Starting from Eq.~(\ref{psi-1}) we write $|\psi(z,t)|^4$ as the
product of four intergals. Performing the standard Gaussian average of
the expression $\exp[{\rm
  i}(\theta(z_1)-\theta(z_2)+\theta(z_3)-\theta(z_4))]$ then yields
\begin{equation}
S(t) + 1 = \frac{\theta_0}{2 \pi \tit} \int \int {\rm d}\tix {\rm
  d}\tiy \, {\rm e}^{-{\rm i}\theta_0 \tix \tiy/\tit}\, {\rm
  e}^{-\theta_0^2 K(\tix, \tiy)}\;.
\label{S}
\end{equation}
The function $K(\tix ,\tiy )$ is defined by
\begin{equation}
K(\tix ,\tiy ) = 2 - 2f(\tix ) - 2 f(\tiy ) + f(\tix +\tiy ) + f(\tix -\tiy )\;,
\end{equation}
where $f(\tix)$ is the correlation function introduced in Eq.~(\ref{disorder-correlator}).

For $\theta_0 \ll 1$, $\exp(-\theta_0^2 K(\tix ,\tiy ))$ may be
expanded as $1--\theta_0^2 K(\tix ,\tiy )$. Then $S(t)$ is small,
given by
\begin{equation}
S(t) = 2 \theta_0^2\left[1 - \frac{1}{\sqrt{2\pi}} \int {\rm d} s\,(\sin s^2 + \cos s^2)  f(s\sqrt{4 \tit /\theta_0})\right]\;,
\end{equation}
and saturates at the value $2 \theta_0^2$ for $\tit \gg
\theta_0$. This small $\theta_0$ regime was studied in
Ref.~\cite{clement}. We turn to the more interesting case, $\theta_0
\gg 1$, when caustics and related large interference effects can be
observed. For this case no expansion in $\theta_0$ is possible and the
analytical treatment becomes tedious \cite{jakeman}. It is possible,
however, to identify three distinct contributions to the integral in
Eq.~(\ref{S}), each of which dominates at the appropriate time, and
thus to derive an approximate expression:
\begin{eqnarray}
S(t) + 1 &\approx& \frac{2}{\pi} \int_0^\infty \frac{{\rm d}s}{s} \sin(\theta_0 s/\tit) \,{\rm e}^{-\frac{1}{2}\gamma \theta_0^2 s^2} \nonumber\\
&&-\frac{2}{\pi} \gamma \int_0^\infty {\rm d}s \, s \sin(s/\tit) \ln(s/\theta_0)\, {\rm e}^{-\frac{1}{2}\gamma s^2} \nonumber \\
&&+ 2\left[ 1 - {\rm erf}\left(\frac{1}{2\sqrt{\beta} \tit}\right)\right]\;,
\label{S1}
\end{eqnarray}
where $\beta = |\partial^2_{\tix}  f( \tix )|_{\tix {= }0} | $ and
$\gamma = |\partial^4_{\tix} f(\tix)|_{\tix{=}0}|$. For the speckle
pattern these numbers are $\beta = 1/3$ and $\gamma = 2/45$. The three
terms in Eq.~(\ref{S1}) make the dominant contribution to the
scintillation index at different times. For short times, $\tit \ll 1$,
the first term dominates and approaches $1$ as $\tit \to 0$, so that
$S(t)$ drops to zero. The last term dominates in the opposite limit,
$\tit \gg 1$. It approaches $2$, thus yielding the expected saturation
value $S(t) =1$. The most interesting term, however, is the second
one. It dominates for intermediate times, when $\tit \sim 1$, and it
is proportional to $\ln \theta_0 $, signalling the appearance at this
time of caustics and the associated large density fluctuations.

\begin{figure}
\includegraphics[scale=0.45]{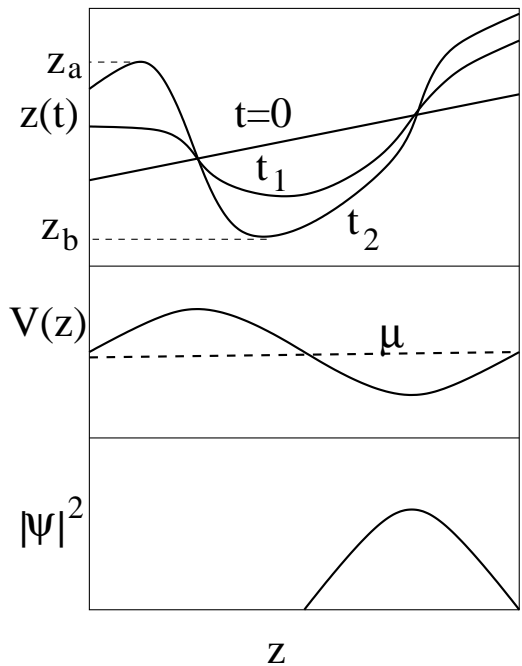}
\caption{Schematic illustration of the effect of condensate
  fragmentation before expansion on caustic formation after expansion.
  Middle panel: potential $V(z)$ and chemical potential $\mu$ before
  expansion. Lower panel: condensate density $|\psi|^2$ before
  expansion. Upper panel: position $z(t)$ reached by atoms after
  expansion as a function of starting point $z$, for three different
  expansion times, $t=0,t_1$ and $t_2$, with $0<t_1<t_2$. Caustics appear at turning points of $z(t)$ vs
  $z$, marked here for $t_2$ at $z_a$ and $z_b$. For the value of $\mu$ illustrated, the initial density is zero
  at points from which caustics would otherwise develop.} \label{fig2}
\end{figure}

Let us emphasize that the two basic requirements for our treatment
are $\hbar \omega_\perp \ll \mu$ and $R_0 \gg a_\perp$. The first
inequality  ensures that the size and energy of the condensate,
while in equilibrium in the trap, is dominated by the interactions
i.e. by the nonlinear term in the Gross-Pitaevskii equation. The
second inequality is required for the validity of the
two-stage-expansion scenario as well as of the ray picture on
which the physics of caustics rests. To this point our treatment
of density modulations in a BEC after expansion has been
restricted to systems in which the relative amplitude of initial
density fluctuations is small, which is the case when these are
produced by a potential with amplitude $V_0 \ll \mu$. It is in
fact straightforward to generalise the discussion, to allow for an
arbitrary value of $V_0$.
A potential with amplitude $V_0 \gtrsim \mu$ automatically
implies impressed phase fluctuations with amplitude $\theta_0 \gg
1$. That in turn justifies a stationary phase treatment of caustic
formation, and according to this treatment, the atoms that form
caustics come from short segments of the condensate, which have
width proportional to $\theta_0^{-1/3}$ before the second stage of
the expansion. Moreover, the density in these segments remains
approximately constant during the first stage of expansion,
provided $R_0 \gg a_\perp$. To obtain the final relative density
after expansion under these conditions, the value of
$|\psi(z,t)|^2$ calculated from Eq.~(\ref{psi-1}) should simply be
multiplied by the initial relative density $|\psi(z',0)|^2$ at a
point $z'$ determined from
\begin{equation}
z'=z - \frac{\hbar t}{m} \partial_{z'} \theta(z')\;.
\end{equation}
The consequences of this are most significant for $V_0 \gtrsim \mu$,
when the condensate before expansion is fragmented, having initial
relative density $|\psi(z',0)|^2 = 0$ for some values of $z'$. In
particular, it turns out that caustics are  completely suppressed for
$V_0 \gg \mu$, because in this limit the initial density is zero at
all points $z'$ for which there is focussing of the emerging rays. To
show this, note that focussing at time $t$ of atoms from $z'$ occurs
if
\begin{equation}
\partial^2_{z'} \theta(z') = - \frac{m}{\hbar t}\;,
\end{equation}
that is to say, $\partial^2_{z'} \theta(z')$ must be negative.
On the other hand, for $V_0 \gg \mu$, the condensate before expansion
occupies the neighbourhood of minima of the potential $V(x)$. In these
regions $\partial^2_{z'} V(z')$ and hence $\partial^2_{z'} \theta(z')$
are positive, while at the points where $\partial^2_{z'} \theta(z')$
is negative, the initial density  is zero.
These ideas are illustrated schematically in Fig.~\ref{fig2}.

We next comment on a recent experiment \cite{chen} which appears to
satisfy the conditions for caustic formation. In this experiment
$\mu/\hbar \omega_\perp = 5.6$, $R_0 = 15 \mu{\rm m}$, $a_\perp
\lesssim 10 \mu{\rm m}$ and the largest density variations were
observed for $V_0 = 0.5 \mu$, in a time of flight image at $t_{\rm
  ToF}=8{\rm ms}$, which is significantly larger than
$1/\omega_\perp$. This value for $V_0$ corresponds to a phase
amplitude $\theta_0=4.4$ which gives the caustic formation time $t^* =
5{\rm ms}$. This is smaller than but comparable to  the value of
$t_{\rm ToF}$, and so the large density variations in Fig.~4(h) of
Ref.~\cite{chen} can most likely be attributed to caustics. While the
value $\theta_0=4.4$ does not lie deep within the large $\theta_0$
regime, the experiment was not designed for caustic observation. The
results presented above should facilitate the optimisation of such
experiments. We note further that at $V_0 = \mu$, the largest potential amplitude
for which results are reported in Ref.~\cite{chen}, the condensate
before expansion appears to be fragmented (Fig.~4(i) of \cite{chen}) while there are no large
density fluctuations after expansion (Fig.~4(j) of \cite{chen}). Both
features are consistent with the scenario presented in our discussion
of Fig.~\ref{fig2}.

A related experiment has been described in Ref.~\cite{clement}
together with a theoretical discussion appropriate for small
$\theta_0$. Some of the parameters for this experiment take similar
values to those of Ref.~\cite{chen}: $\mu/\hbar \omega_\perp = 6.8$ and
$V_0 = 0.41 \mu$, giving $\theta_0 = 4.3$. However, the length scales
$R_0 = 1.7 \mu {\rm m}$ and $a_\perp = 1.5 \mu {\rm m}$ are
about an order of magnitude smaller than the corresponding ones in
\cite{chen}. These lengths are also much smaller
than the reported resolution ($8.5 \mu{\rm m}$) of the imaging system
and of the density fluctuations after expansion that are illustrated
in Fig.~1 of Ref.~\cite{clement}. We note in addition that the
time of flight used in \cite{clement} (in the range $5$-$17{\rm ms}$)
is about an order of magnitude larger than the value $t^* = 1{\rm ms}$
we calculate from experimental parameters. Therefore caustics in this
experiment should form and decay before time of flight images are
made, and should have a smaller spacing than the resolution of the
imaging system. For these reasons we do not
expect the theory we have presented to apply directly to the
measurements of Ref.~\cite{clement}.

\section{Strictly one dimension: many body correlations}
\label{s1d}

In this section we discuss the possibility of caustic formation, in a
system of interacting bosons, beyond the mean field approach. We
assume here that $\hbar \omega_\perp$ is much larger than the
characteristic interaction energy so that, with respect to transverse
motion, all $N$ atoms in the trap reside in the ground state
$\chi_0(\rho)$ of the harmonic oscillator, forming a strictly
one-dimensional system. There is no confining potential in the axial
direction, in which particles move on an interval $L$ with periodic
boundary conditions. The axial motion is controlled by an effective
one-dimensional Hamiltonian -- the Lieb-Liniger model
\cite{lieb,olshanii} . The chemical potential of the system, in
equilibrium in the trap, differs from $\hbar \omega_\perp$ only by a
small, interaction induced correction. This correction is crucial for
the ground state properties of the gas. However, when the gas is
released from the trap, the radial expansion will be governed not by
the interaction, as happened for the many channel case ($\mu \gg \hbar
\omega_\perp$) considered in Sec. \ref{q1d}, but by the zero-point
energy associated with radial motion. Thus, for a strictly
one-dimensional system it is not possible to generate a large phase
imprint from small initial density modulation during the radial
expansion. Therefore, in this section as a
means of impressing a phase we employ the second possibility, namely a
short potential pulse, as mentioned in Sec. \ref{introduction}. Phase
imprinting of this kind has been proposed \cite{dobrek} and used
extensively to generate vortices in two and three dimensional
condensates. The
mechanism works as follows: starting at time $-\tau$ we apply to the
system, in its ground state, a short potential pulse, of duration $\tau$ and of a prescribed spatial profile
\begin{equation}
V(z,t)=- \frac{\hbar}{\tau} \theta(z)\;, \qquad -\tau < t < 0
\end{equation}
where $\theta$ can be a deterministic or a random function of $z$. If
the time interval $\tau$ is shorter than the characteristic times of
the system, then at the time instant $t=0$ the axial part of the
many-body wavefunction will acquire a phase
\begin{equation}
\Psi(z_1, \ldots z_N;t{=}0) = \exp\left[{\rm i}\sum_{j=1}^N \theta(z_j)\right] \Phi_0(z_1, \ldots z_N)
\label{initial-1d}
\end{equation}
where $ \Phi_0(z_1, \ldots z_N)$ is the ground state wavefunction,
prior to the action of the pulse, and is normalised to unity.  (The
complete wavefunction for the system in three dimensions of course
also includes the radial factor $\prod_{j=1}^N\chi_0(\rho_j)$.) At
time $t=0$, just after this phase has been impressed, the trapping
potential is switched off and the gas undergoes radial expansion. The
initial Gaussian function, $\chi_0(\rho)$, will spread with time,
retaining its Gaussian shape, and the density of the gas will evolve
accordingly.  Therefore we shall neglect interactions during the
expansion, so that the gas is assumed to undergo free evolution and
the $z$-dependent part of the many-body wavefunction evolves according
to
\begin{equation}
{\rm i} \hbar \frac{\partial \Psi(z_1, \ldots z_N; t)}{\partial t} = \left[ \sum_{j=1}^N\left( - \frac{\hbar^2}{2m} \frac{\partial^2}{\partial z_j^2}\right)\right] \Psi(z_1, \ldots z_N;t)\;,
\end{equation}
which is to be solved with the initial condition given in
Eq.~(\ref{initial-1d}). This initial function contains all the
information on the interacting groundstate, prior to the expansion.

We are interested in the one-dimensional, $z$-dependent part of the particle density
\begin{equation}
n_1(z,t) = N \int_0^L \left| \Psi(z,z_2 \ldots z_N;t)\right|^2 {\rm d}z_2 \ldots {\rm d}z_N \equiv \frac{N}{L}F(z,t)\;.
\end{equation}
The actual, three-dimensional density $n(\rho,z,t)$ is obtained by
multiplying $n_1(z,t)$ by the radial factor $|\chi(\rho,t)|^2$ --  the
spreading Gaussian. Initially, $F(z,t{=}0) = 1$ but, as the expansion
proceeds, $F(z,t)$ develops modulations in $z$, due to the initially
impressed phase. In second quantised form
\begin{equation}
F(z,t) = \frac{L}{N} \langle \Psi | \hat{\psi}^\dagger(z,t) \hat{\psi}(z,t) | \Psi \rangle
\label{1d-density}
\end{equation}
where $|\Psi \rangle$ is the initial state vector, defined in position
representation in Eq.~(\ref{initial-1d}), while
$\hat{\psi}^\dagger(z,t)$ and ${\psi(z,t)}$ are the free field operators
\begin{equation}
\hat{\psi}^\dagger(z,t) = \frac{1}{\sqrt{L}}\sum_{k} {\rm e}^{{\rm i} \frac{\hbar k^2}{2m}t - {\rm i}kz}a_k^\dagger\;
\label{expansion}
\end{equation}
with creation and annihilation operators $a^\dagger_k$ and
$a^{\phantom{\dagger}}_k$ that satisfy the commutation relation $[a^{\phantom{\dagger}}_q, a^\dagger_k]=\delta_{k,q}$.

Using Eq.~(\ref{expansion}) and the relation
\begin{equation}
\langle  \Psi | \hat{\psi}^\dagger(x) \hat{\psi}(y) | \Psi \rangle =
{\rm e}^{{\rm i} (\theta(y) - \theta(x)) } \langle  \Phi_0 |
\hat{\psi}^\dagger(x) \hat{\psi}(y) |  \Phi_0\rangle \;,
\end{equation}
Eq.~(\ref{1d-density}) can be written as
\begin{equation}
F(z,t) = \int \frac{{\rm d}p}{2\pi} |G_p(z,t)|^2 n(p)\;,
\label{1d-density-result}
\end{equation}
where $n(p)$ is the momentum distribution function
\begin{equation}
n(p) = \frac{L}{N} \int {\rm d} z \langle \hat{\psi}^\dagger(z)\hat{\psi}^{\phantom{\dagger}}(0) \rangle {\rm e}^{-{\rm i}pz}
\end{equation}
normalised so that
\begin{equation}
\int  \frac{{\rm d}p}{2\pi}  n(p) = 1\;.
\end{equation}
The function $G_p(z,t)$ is
\begin{equation}
G_p(z,t) = \sqrt{\frac{m}{2\pi {\rm i} \hbar t}} \int {\rm d}\zeta
\exp\left[\frac{{\rm i} m}{2 \hbar t}(z-\zeta)^2 +{\rm i}
  \theta(\zeta) - {\rm i} {p} \zeta\right]\;.
\label{G}
\end{equation}
In the mean field approach $n(p) = 2\pi \delta(p)$ and the expression
for $G_p(z,t)$ reduces to that for $\psi(z,t)$ given in
Eq.~(\ref{psi-1}) of Sec.~\ref{q1d}. Such behaviour corresponds to the
non-interacting limit of the Leib-Liniger gas which, under radial
expansion, will exhibit large density variations and caustics, as
discussed in Sec.~\ref{q1d}. Below we show that interactions inhibit
caustic formation, and derive a condition for caustics to survive in
the presence of interactions.

The treatment of the integral in Eq.~(\ref{G}) is along the same lines
as in Sec \ref{q1d}.  Caustics originate from rays emerging from
points $\zeta_n$ at which the second derivative of the phase in
Eq.~(\ref{G}) vanishes. Since this derivative does not contain $p$, we
have the familiar condition for the singular points, $\partial^2_\zeta
\theta(\zeta) = -m/\hbar t$ at $\zeta=\zeta_n$. The first derivative,
however, now contains $p$ so that the analogue of Eq.~(\ref{rays}),
for an arbitrary function $\theta(\zeta)$ and keeping all physical
dimensions, is
\begin{equation}
z = \zeta + \frac{\hbar t}{m}\left( \partial_\zeta \theta(\zeta) - p \right)\;.
\end{equation}
This is essentially the definition of a ray, emerging from a point
$\zeta$, and having a particular value of $p$. The dependence on $p$
implies that the rays emerging from the same singular point $\zeta_n$, but
having different values of $p$, will not arrive at the point $z$ at
the same time. Thus focussing, which is the essence of the phenomenon
of caustics, will be suppressed and caustics will get washed out.  The
condition for the existence of caustics follows from the integral in
Eq.~(\ref{G}). Assuming that the function $\theta(\zeta)$ has a
characteristic amplitude $\theta_0$ and scale of variation $R_0$, and
expanding the phase in Eq.~(\ref{G}) near a singular point $\zeta_n$,
one obtains a contribution $C \theta_0 R_0^{-3} \eta^3 -
{p} \eta$ where $\eta = \zeta - \zeta_n$ and $C$ is a
constant of order $1$. Caustics originate from the $\eta^3$-term and
the relevant range of integration is $\eta \lesssim R_0
\theta_0^{-1/3}$. Therefore, if the contributing values of $p$,
defined by the range over which $n(p)$ is significant, are such that
$R_0 p  \theta_0^{-1/3} \ll 1$, then caustics will
survive. Since most of the weight of the momentum distribution $n(p)$
is concentrated in the interval $p\lesssim 1/\xi$
\cite{haldane,astrakharchik}, where $\xi$ is the coherence length for
the Bose gas, we arrive at the condition $\theta_0^{1/3} \xi/R_0 \gg
1$. It should be supplemented by the requirement
that, for a caustic to be formed and visible, there must be
many particles in the region from
which it originates. This condition is $n_1 R_0\theta_0^{-1/3}\gg 1$. Both conditions can be
easily satisfied in the weak interaction limit where $\xi n_1 \gg
1$. The situation is different in the opposite case of strong
interactions, or hard-core bosons. In this case $\xi \approx 1/n_1$,
and so caustics are absent in the
strongly interacting limit.

\section{Two dimensional systems}
\label{2d}

In this section we consider expansion of a quasi-two dimensional
condensate after release from a trap with strong axial and weak radial
confinement, characterised as in Sec~\ref{q1d} by frequencies
$\omega_z$ and $\omega_\perp$ but here with $\omega_z \gg
\omega_\perp$. Again we neglect the weaker confinement, restricting
our discussion to times $t \ll 1/\omega_\perp$ after release of the
trap, and treat a BEC with initial density modulations, which are
converted during expansion into an impressed phase. A difference
between the two-dimensional and one-dimensional geometries is that
caustics in two dimensions consist of lines rather than
points. Another difference is that, within the Thomas-Fermi
approximation, a two-dimensional condensate in a disordered potential
undergoes a percolation transition at a finite critical value of
disorder strength. Because caustic formation is essentially a local
pheomenon, the network of caustic lines does not show any critical behaviour that
reflects this percolation transition, although it does change
with disorder strength.

The distinction between the initial and late phases of expansion is
not as sharp in two-dimensions as it is in one dimension. The reason
for this is that the characteristic density near the center of the
trap decreases with time $t$ as $t^{-1}$ in two dimensions and as
$t^{-2}$ in one dimension, with the result that the impressed phase
grows logarithmically at long times in two dimensions, but reaches a
limiting
value in one dimension. We neglect this logarithmic growth and take
the impressed phase to have a definite value
\begin{equation}
\theta({\bf r}_\perp) \sim \frac{V({\bf r}_\perp)}{\hbar \omega_z}\;
\end{equation}
at times large compared to $1/\omega_z$,
with characteristic amplitude $\theta_0$ and length scale $1/k_0$.
In this approximation the wavefunction during the second phase of
expansion can be factorised as
\begin{equation}
\Psi({\bf r}_\perp,z,t) = \Phi({\bf r}_\perp,t)\psi(z,t)\;.
\end{equation}
For a two-dimensional system the axial part is given by a scaling function \cite{pitaevskii}
and our interest is in the evolution of the planar part,
$\Phi({\bf r}_\perp,t)$. In analogy with Eq.~(\ref{psi-1}), it is given at
late times by
\begin{equation}
\Phi({\bf r}_\perp,t) = \frac{m}{2\pi {\rm i} \hbar t} \int {\rm
  d}^2{\bf r}_\perp^\prime \exp\left[\frac{{\rm i}m}{2\hbar t}|{\bf
    r}_\perp - {\bf r}_\perp^\prime|^2 +{\rm i} \theta({\bf
    r}_\perp^\prime)   \right] \;.
\label{planar-phi}
\end{equation}

As in quasi-one dimensional systems, caustics are formed for $\theta_0
\gg 1$ at values of the scaled time $\tit \gtrsim 1$, and in this
regime Eq.~(\ref{planar-phi}) can be evaluated using the stationary
phase method. The saddle points in ${\bf r}_\perp^\prime$ are
the solutions to
\begin{equation}
\frac{m}{\hbar t} ({\bf r}_\perp^\prime - {\bf r}_\perp) +
{\bf \nabla}_{{\bf r}_\perp^\prime} \theta({\bf r}_\perp^\prime) =
0\;.
\label{2dsaddle}
\end{equation}
In the leading approximation, one such saddle point, at ${\bf r}_\perp^\prime = {\bf r}_\perp^*$,
makes a contribution to $\Phi({\bf r}_\perp,t)$ of
modulus $[{\rm det} M({\bf r}_\perp^*)]^{-1/2}$, where
\begin{equation}
M({\bf r}_\perp^\prime)=\left(\begin{array}{cc}
\frac{m}{\hbar t} + \partial^2_{x^\prime} \theta({\bf r}_\perp^\prime) &
\partial_{x^\prime} \partial_{y^\prime} \theta({\bf r}_\perp^\prime)\\
&\\
\partial_{y^\prime} \partial_{x^\prime} \theta({\bf r}_\perp^\prime) &
\frac{m}{\hbar t} + \partial^2_{y^\prime} \theta({\bf r}_\perp^\prime)
\end{array}\right)\;.
\end{equation}
Caustics stem from those saddle points at which ${\rm det} M({\bf
  r}_\perp^*) = 0$.  Atoms in the expanding condensate originating
from these points are focussed in such a way that the density
$|\Phi({\bf r}_\perp,t)|^2$ is divergent within
this approximation, which is equivalent to geometrical optics. To find the
density in the vicinity of caustics, it would be necessary to take
into account higher derivatives of $\theta({\bf r}_\perp)$ when calculating the integral
in Eq.~(\ref{planar-phi}). We do not do this, restricting ourselves instead to a
discussion of the positions of caustics. The condition  ${\rm
  det} M({\bf  r}_\perp^*) = 0$ defines a set of lines in the atomic
cloud after the initial phase of expansion, which give rise to
caustics. Both the number and the shape of these lines in the initial plane
depend on the time at which density in the expanding cloud is to be
measured, but they have fixed limits for $\tit \gg 1$. Atoms starting from
points on these lines travel with velocity $(\hbar/m) \nabla
\theta$, and reach points in the cloud at time $t$ with coordinates
${\bf r}_\perp$ given by the solutions to Eq.~(\ref{2dsaddle}). In this way lines in
the initial plane are mapped to  moving lines of high density in the expanding
cloud.

To illustrate these general ideas, consider the example of a periodic
impressed phase
\begin{equation}
\theta({\bf r}_\perp) = \theta_0 ( \cos k_0 x + \cos k_0 y)\;.
\label{2d-periodic}
\end{equation}
In this case the condition ${\rm det} M({\bf
  r}_\perp^*) = 0$ yields
\begin{equation}
(\tit^{-1} - \cos k_0x^*)(\tit^{-1} - \cos k_0 y^*) = 0\;,
\end{equation}
defining two sets of parallel lines in the initial
plane, $x^* = k_0^{-1} \cos^{-1}(1/ \tit)$ and  $y^* =
k_0^{-1}\cos^{-1}(1/\tit)$. Caustics derive from these, forming two similar sets of lines
in the final plane, $x = x^* - \tit \sin k_0 x^*$ and $y = y^* - \tit
\sin k_0 y^*$ respectively.

\begin{figure}
\includegraphics[scale=0.38]{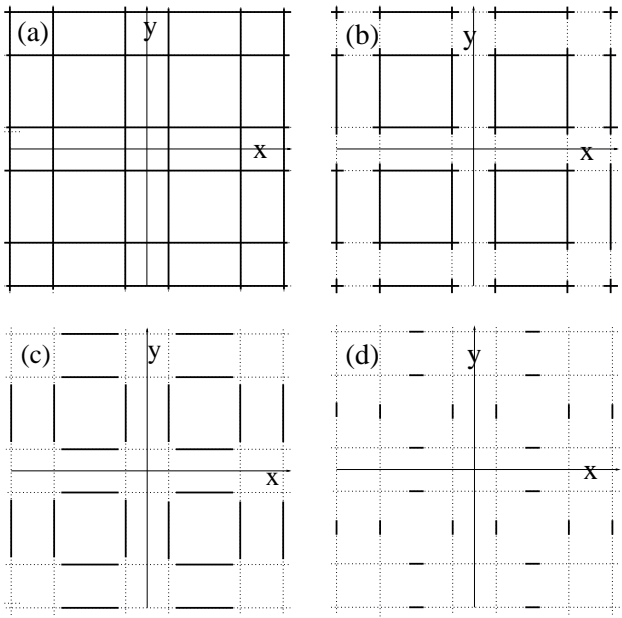}
\caption{Effect of decreasing
  $\mu$ in the condensate before expansion, on caustic formation in two
  dimensions, illustrated for a periodic
  potential $V({\bf r}_\perp)$. Each of the four panels
  shows the initial plane. Caustics evolve from lines with  ${\rm det} M({\bf
  r}_\perp) = 0$ and  $|\Phi({\bf  r}_\perp,0)|^2 > 0$. These are
marked in bold, while lines with ${\rm det} M({\bf
  r}_\perp) = 0$ and  $|\Phi({\bf  r}_\perp,0)|^2 =0$ are shown dotted.
  Panel (a): large $\mu$, $|\Phi({\bf  r}_\perp,0)|^2>0$ everywhere;
  (b) on reducing $\mu$, the  condensate develops small holes but
  percolates, while caustics become fragmented;
  (c) reducing $\mu$ further, the condensate reaches the percolation
  threshold, but caustics show no signature of this;
(d) reducing $\mu$ still further, caustics shrink and eventually disappear.} \label{fig3}
\end{figure}

The consequences in two dimensions of a potential $V({\bf r}_\perp)$ strong enough to
generate significant density modulations before expansion can be
discussed for large $\theta_0$ using the same approach as in
Sec.~\ref{q1d}. In this way we find that two conditions must be
satisfied in order that a line segment in the initial plane will give
rise to a caustic after expansion: it is necessary, first, that ${\rm det} M({\bf
  r}_\perp) = 0$ and, second, that the initial relative density $|\Phi({\bf
  r}_\perp,0)|^2$ is non-zero on the line. In consequence, as the potential strength
is increased, or the chemical potential is reduced, caustic lines
first develop breaks, and then disappear altogether. Such an evolution
with decreasing $\mu$ of the lines in the inital plane that generate
caustics is illustrated in Fig~\ref{fig3} for the periodic potential
that underlies Eq.~(\ref{2d-periodic}).

The theory of caustic formation resulting from a random phase $\theta({\bf
  r}_\perp)$ is analogous to the treatment of the phase screen
problem, which has been studied extensively for two-dimensional
systems in the context of optics \cite{berry}. In particular, the
morphology of caustic lines for the random case is discussed in \cite{berry2}.

\section{Fermions}
\label{fermions}

It is interesting to ask about problems similar to the ones we have discussed, but with
fermions in place of bosons. To be specific, consider expansion of a
strictly one dimensional system with an imprinted phase, as in
Sec.~\ref{s1d}, but for non-interacting fermions rather than
interacting bosons. Restricting our attention to the expectation value
of the density as a function of position and time, the effects of
particle statistics enter the main result,
Eq.~(\ref{1d-density-result}), only through the momentum distribution
of particles before expansion. The criterion for formation of caustics
in the expanding Fermi gas is therefore the same as for the
Lieb-Liniger gas, but with the Fermi wavelength $\lambda_{\rm F}$ taking
the place of the coherence length $\xi$ in the expressions given in
Sec.~\ref{s1d}. Hence caustics are absent from the
Fermi gas, for the same reason as in the
Bose gas with hard-core interactions. There are
nevertheless differences between non-interacting fermions and hard-core
bosons. They stem from the
Fermi surface discontinuity in the momentum distribution. Within the
approximations of geometrical optics, this discontinuity leads to
sharp peaks in the {\it derivative} of the relative density $F(z,t)$
with respect to position $z$ or time $t$. To show this, we note from
Eq.~(\ref{G}) that
\begin{equation}
\partial_z |G_p(z,t)|^2 = \frac{m}{\hbar t} \partial_p  |G_p(z,t)|^2\;.
\end{equation}
With Eq.~(\ref{1d-density-result}) this yields
\begin{eqnarray}
\partial_z F(z,t) &=& - \frac{m}{\hbar t} \int \frac{{\rm d}p}{2 \pi} |G_p(z,t)|^2
\partial_p n(p)\nonumber\\
&=& \frac{m}{2\hbar p_{\rm F}t}\left[ |G_{p_{\rm F}}(z,t)|^2 - |G_{-p_{\rm
      F}} (z,t)|^2  \right]\,,
\end{eqnarray}
where $p_{\rm F}$ is the Fermi wavevector. The behaviour of $G_p(z,t)$
has been analysed in Sec.~\ref{q1d}: the value of $p$ influences the
position of caustics
but not their formation. The Fermi gas therefore shows the same extrema
in the derivative of the density as are found for a BEC in the density
itself.

\section{Summary}
\label{summary}

We have discussed free motion of one and two dimensional atomic gases
with an initial impressed phase that varies periodically or randomly
as a function of position. Gradients of this phase represent initial velocities
and lead to density variations that grow with time. The characteristic
amplitude $\theta_0$ and size $R_0$ of spatial variations of this
phase are key parameters. The limit $\theta_0 \gg 1$ is both the most
interesting regime, because density maxima are largest, and a
tractable one theoretically, because a treatment analogous to
geometrical optics provides the leading approximation. The evolution
of atomic density fluctuations with time has close links to problems
in optics involving caustic formation. In the context of atomic gases,
caustics are maxima of density, near points in one dimensional systems
or along lines in the two dimensional case. For atoms
of mass $m$ they form at a characteristic time $t^* = mR_0^2/\hbar \theta_0$
and at longer times $t$ the density within a caustic decays as
$t^{-1}$. Since caustics originate from small regions of the initial
atomic cloud, variations in the initial density simply modulate the
density on caustics. In particular, caustic formation is suppressed during expansion
of a fragmented condensate if the initial density is zero at points
that would otherwise be the origin for caustics.

We have argued that a recent experiment \cite{chen} in which large density modulations are
observed in an elongated BEC after release from a disordered potential
should be understood in terms of caustic formation. For the future it
would be of interest to design experiments with larger values of
$\theta_0$ for both one and two dimensional systems.


\acknowledgments
This work was supported in part by the Royal Society through the award of a visiting fellowship to B~S. and by
EPSRC under Grant No. EP/D050752/1.


\end{document}